\documentclass[journal]{IEEEtran}

\usepackage{amsmath,graphicx,hyperref,amssymb,tikz,pgfplots,cite,comment,algorithm,algpseudocode,subcaption}

\renewcommand{\a}{\mathbf{a}}

\newcommand{\h}{\mathbf{h}}

\renewcommand{\r}{\mathbf{r}}
\newcommand{\s}{\mathbf{s}}

\newcommand{\0}{\mathbf{0}}

\newcommand{\A}{\mathbf{A}}
\newcommand{\B}{\mathbf{B}}
\newcommand{\C}{\mathbf{C}}

\newcommand{\E}{\mathbf{E}}

\newcommand{\G}{\mathbf{G}}
\renewcommand{\H}{\mathbf{H}}

\renewcommand{\P}{\mathbf{P}}

\newcommand{\Y}{\mathbf{Y}}

\newcommand{\phib}{{\boldsymbol{\phi}}}

\newcommand{\setC}{\mathcal{C}}

\newcommand{\setN}{\mathcal{N}}

\newcommand{\Compl}{\mbox{$\mathbb{C}$}}
\newcommand{\Real}{\mbox{$\mathbb{R}$}}

\newcommand{\Exp}{\mathbb{E}}
\newcommand{\herm}{\mathrm{H}}

\newcommand{\minimize}{\mathrm{minimize}}

\renewcommand{\Re}{\mathrm{Re}}

\newcommand{\tr}{\mathrm{tr}}
\newcommand{\tran}{\mathrm{T}}

\definecolor{oulu_blue}{HTML}{23408F}
\definecolor{oulu_green}{HTML}{39B54A}

\definecolor{red}{rgb}{1,0,0}
\definecolor{red_magenta}{rgb}{1,0,0.5}
\definecolor{magenta}{rgb}{1,0,1}
\definecolor{blue_magenta}{rgb}{0.5,0,1}
\definecolor{blue}{rgb}{0,0,1}
\definecolor{blue_cyan}{rgb}{0,0.5,1}
\definecolor{cyan}{rgb}{0,1,1}
\definecolor{green_cyan}{rgb}{0,1,0.5}
\definecolor{green}{rgb}{0,1,0}
\definecolor{green_yellow}{rgb}{0.5,1,0}
\definecolor{yellow}{rgb}{1,1,0}
\definecolor{red_yellow}{rgb}{1,0.5,0}

\ifCLASSINFOpdf
\else
   \usepackage[dvips]{graphicx}
\fi
\usepackage{url}

\usepackage[figurename=Fig., labelsep=period, font=small]{caption}

\usepackage{graphicx}

\usepackage{titlesec}
\let\subparagraph\relax
\titlespacing{\section}{0pt}{6pt plus 2pt minus 1pt}{4pt plus 1pt minus 1pt} 
\titlespacing{\subsection}{0pt}{5pt plus 2pt minus 1pt}{3pt plus 1pt minus 1pt} 
\definecolor{myBlue}{RGB}{31,119,180}
\definecolor{myOrange}{RGB}{255,127,14}
\definecolor{myGreen}{RGB}{44,160,44}
\definecolor{myPurple}{RGB}{148,103,189}
\definecolor{myRed}{RGB}{214,39,40}

\begin{document}

\title{ISTA-Based Joint Dictionary Learning and Channel Estimation for XL-MIMO Systems}

\author{Arttu Arjas and Italo Atzeni
\thanks{The authors are with the Centre for Wireless Communications, University of Oulu, Oulu, Finland (e-mail: \{arttu.arjas, italo.atzeni\}@oulu.fi).}
\thanks{This work was supported by the Research Council of Finland (336449 Profi6, 348396 HIGH-6G, and 369116 6G~Flagship).}}

\maketitle

\begin{abstract}
Channel estimation in extra-large multiple-input multiple-output systems is challenging due to near-field propagation, where the array response depends on both the angle and distance of the propagation paths. Existing near-field channel estimation methods typically rely either on fixed angle-distance grids, which suffer from grid mismatch effects, or on multi-stage refinement procedures with increased computational complexity. To address these limitations, this paper proposes the \textit{dictionary-learning iterative soft-thresholding algorithm (DL-ISTA)}, a method for joint near-field dictionary learning and channel estimation based on the iterative soft-thresholding algorithm. The proposed method jointly estimates the sparse channel coefficients and the continuous angle-distance parameters through alternating optimization, thereby avoiding discretization errors associated with fixed grids. To promote robust convergence, the angle-distance parameters are initialized using Sobol sequences, which provide uniform coverage of the parameter space. Numerical results show that DL-ISTA outperforms a baseline with comparable computational complexity and attains comparable or better accuracy than a substantially more complex benchmark.
\end{abstract}

\begin{IEEEkeywords}
Channel estimation, dictionary learning, ISTA, near-field communications, XL-MIMO.
\end{IEEEkeywords}

\section{Introduction}

Accurate channel estimation is crucial in multiple-input multiple-output (MIMO) systems to enable efficient directional transmission through beamforming \cite{Bjo17}. This task becomes especially important in extra-large MIMO (XL-MIMO), where large antenna arrays and high carrier frequencies give rise to near-field propagation \cite{Atz25}. In this regime, the waves seen at the base station (BS) must be modeled as spherical, making the array response dependent on both angle and distance \cite{cui2022near}.

In XL-MIMO settings, the channels typically exhibit sparsity in the polar domain, with the signal propagating through only a few dominant paths. By exploiting this sparsity, compressive-sensing techniques can significantly reduce the channel estimation overhead \cite{berger2010application}. A key challenge in this context is constructing a dictionary that accurately represents the continuous angle-distance parameter space. In near-field channel estimation, this dictionary is parameterized by the steering angles and distances of the received signal. Since these parameters are unknown, a common approach is to discretize them using a predefined grid. In \cite{cui2022channel}, a low-coherence dictionary was constructed from such a grid and used to develop the polar-domain orthogonal matching pursuit (P-OMP) channel estimation method. However, when the actual angle-distance pairs do not lie on the grid, the estimation accuracy degrades due to basis mismatch. To mitigate this issue, a gradient-descent refinement step, referred to as polar-domain simultaneous iterative
gridless weighted (P-SIGW) method, was employed to improve the angle and distance estimates \cite{cui2022channel}. Similarly, \cite{zhang2023near,lu2023near} proposed two-stage approaches with refinement steps to reduce the performance loss caused by grid mismatch effects. The method in \cite{ye2024gan} addressed this problem from a data-driven perspective, but requires training data and incurs the associated training overhead. More recently, our prior work \cite{arjas2026cross} proposed a cross-predictive sparse Bayesian learning (CP-SBL) framework for near-field channel estimation. Nevertheless, these approaches still rely either on discretized parameter representations and multi-stage procedures, or on additional training data, motivating the need for methods that operate directly in the continuous parameter space.

In this paper, we propose a novel near-field channel estimation method based on joint dictionary learning and the iterative soft-thresholding algorithm (ISTA) \cite{daubechies2004iterative}. The proposed \textit{dictionary-learning ISTA (DL-ISTA)} jointly estimates the sparse channel coefficients and the continuous angle-distance parameters via alternating optimization, thereby avoiding discretization errors associated with fixed angle-distance grids. Since the resulting optimization problem is highly non-convex, the initialization of the angle-distance parameters plays a crucial role in the performance. To promote robust convergence, we initialize the angle-distance pairs using Sobol sequences \cite{sobol1967distribution}, which provide uniform coverage of the parameter space. Unlike existing two-stage approaches, such as P-OMP+P-SIGW, DL-ISTA performs channel estimation and parameter refinement within a single optimization framework. Furthermore, it achieves similar or better performance than CP-SBL with approximately one order of magnitude lower computational complexity. The main contributions are summarized as follows:
\begin{itemize}
    \item We propose DL-ISTA, a method for joint near-field dictionary learning and channel estimation based on ISTA.
    \item We employ Sobol sequences to initialize the continuous angle-distance parameters, promoting convergence to a favorable local optimum.
    \item We demonstrate that DL-ISTA consistently achieves the best overall performance, surpassing P-OMP+P-SIGW while matching or exceeding the performance of CP-SBL with a substantially lower computational cost.
\end{itemize}

\textit{\textbf{Notation.}} Vectors and matrices are denoted by bold lowercase and uppercase letters, respectively. The transpose and Hermitian transpose operators are denoted by $(\cdot)^\tran$ and $(\cdot)^\herm$, respectively. The Frobenius inner product is defined as $\langle \B, \C \rangle = \tr(\B^\herm \C)$. The elementwise matrix $\ell_{1}$-norm is denoted as $\|\B\|_{1,1} = \sum_{m,n}|B_{mn}|$. The complex soft-thresholding operator is given by $\mathcal{S}_{\gamma}(b) = \max(|b| - \gamma, 0)\frac{b}{|b|}$, with $\mathcal{S}_{\gamma}(0) = 0$. Lastly, $\Pi_{[a,b]}(\cdot)$ denotes projection onto the interval $[a,b]$.

\section{System Model}

In this section, we describe the system and channels models, as well as the sparse channel representation. Consider an XL-MIMO system where a BS with $M$ antennas serves $K$ single-antenna UEs. The channels of the UEs are concatenated into the channel matrix $\H = [\h_1, \dots, \h_K] \in \Compl^{M \times K}$.~In~each coherence block, the UEs simultaneously transmit $N$~pilot~symbols, collected in the pilot matrix $\P \in \Compl^{K \times N}$, to~enable uplink channel estimation. The resulting received pilot signal is
\begin{equation} \label{eq:systemmodel_0}
\Y = \sqrt{\rho} \H \P + \E \in \Compl^{M \times N},
\end{equation}
where $\rho > 0$ is the transmit power and $\E \in \Compl^{M \times N}$ is a matrix of additive white Gaussian noise (AWGN) with variance $\sigma^2$.

\subsection{Channel Model}

In XL-MIMO systems, the large array aperture and high carrier frequency often result in near-field propagation, where the curvature of the impinging wavefront becomes significant and the array response depends on both angle and distance. The boundary of the near-field region is commonly characterized by the Fraunhofer distance $r_{\textrm{Fraun}}=\tfrac{2D^2}{\lambda}$, where $D$ denotes the BS array aperture and $\lambda$ the carrier wavelength. Neglecting the distance dependence of the array response can therefore lead to degraded channel estimation performance.

Considering a uniform linear array (ULA) at the BS and a near-field clustered multipath model, the channel for the $k$-th UE can be expressed as \cite{cui2022channel}, \cite[Ch.~5.6.1]{Bjo24}
\begin{equation} \label{eq:channelmodel}
    \h_k = \sqrt{\frac{M}{L}} \sum_{l=1}^{L} g_{kl} \, \a(\theta_{kl}, r_{kl}) \in \mathbb{C}^{M},
\end{equation}
where $L$ is the number of propagation paths, $g_{kl} \sim \setC \setN (0,1)$ denotes the complex path gain of the $l$-th path, and $\theta_{kl} \in [0,\pi]$ and $r_{kl} > 0$ are the angle or arrival and distance of the $l$-th path, respectively, referenced to the center of the ULA. The vector $\mathbf{a}(\theta, r) \in \Compl^{M}$ represents the ULA near-field steering vector associated with angle $\theta$ and distance $r$, defined as
\begin{equation} \label{eq:steering_vector}
    \mathbf{a}(\theta, r) = \sqrt{\frac{1}{M}} \big[
    e^{-j\frac{2\pi}{\lambda}\left(r_1 - r\right)}, \dots, 
    e^{-j\frac{2\pi}{\lambda}\left(r_{M} - r\right)}
    \big]^{\tran}.
\end{equation}
Letting $\delta_m = \frac{2m - M - 1}{2}$ for $m \in \{1, \dots, M \}$, the distance relative to the $m$-th antenna is given by
\begin{equation}
    r_m = \sqrt{r^2 + (\delta_m d)^2 - 2 r \, \delta_m d \cos \theta},
\end{equation}
where $d$ denotes the inter-antenna spacing of the ULA. While we assume that the BS is equipped with a ULA, the proposed method is applicable to any array geometry.

\subsection{Sparse Channel Representation}

To capture the underlying physical structure of the channel, we parametrize $\H$ in terms of the complex path gains, angles, and distances. We define the angle cosine $\phi = \cos \theta \in [-1, 1]$ and consider a set of $Q$ candidate angle-distance pairs $\{(\phi_q, r_q)\}_{q=1}^Q$, which are used as dictionary. We collect the angles and distances in the vectors $\phib = [\phi_1, \ldots, \phi_Q]^{\tran} \in \Real^Q$ and $\r = [r_1, \ldots, r_Q]^{\tran} \in \Real^Q$, which are used to form the steering matrix $\A(\phib, \r) = [\a(\phi_1, r_1), \dots, \a(\phi_Q, r_Q)] \in \Compl^{M \times Q}$. The corresponding complex path gains are collected in the sparse matrix $\G \in \Compl^{Q \times K}$, where nonzero coefficients correspond to active paths. The channel matrix can then be approximated as $\H \approx \A(\phib, \r)\G$, where equality holds only when the dictionary includes the actual angle-distance pairs.

In this setting, the received signal in \eqref{eq:systemmodel_0} can be expressed~as
\begin{equation} \label{eq:systemmodel}
\Y = \sqrt{\rho}\A(\phib, \r)\G\P + \widetilde{\E},
\end{equation}
where $\widetilde{\E} \in \Compl^{M \times N}$ represents the effective error comprising both AWGN and discretization errors. The goal of this paper is to jointly estimate the steering matrix parameters and the complex path gains.

\section{Proposed DL-ISTA}

In this section, we propose an ISTA-based algorithm for near-field channel estimation, referred to as \textit{dictionary-learning ISTA (DL-ISTA)}, analyze its computational complexity, and discuss the initialization of the angles and distances.

\subsection{Problem Formulation}
We begin by formulating the near-field channel estimation problem as an $\ell_1$-regularized optimization problem, given by
\begin{equation} \label{eq:DL_ISTA_objective}
\underset{\G, \phib, \r}{\minimize}
\; \left\| \A(\phib, \r)\G\P - \Y \right\|_{\mathrm{F}}^2 + \alpha \|\G\|_{1,1},
\end{equation}
where $\alpha > 0$ is a regularization parameter promoting sparsity. The above problem is non-convex due to the coupling between $\G$, $\phib$, and $\r$, making direct joint optimization challenging. To address this challenge, we adopt an alternating minimization approach in which the variables are updated sequentially. In the following, we first derive the update of the sparse matrix $\G$ followed by the updates of $\phib$ and $\r$.

\subsection{Update of the Complex Path Gains}

In this subsection, we write $\A(\phib, \r)$ simply as $\A$ for notational convenience. With $\phib$ and $\r$ fixed, \eqref{eq:DL_ISTA_objective} reduces to
\begin{equation}
\underset{\G}{\minimize}
\; \left\| \A\G\P - \Y \right\|_{\mathrm{F}}^2 + \alpha \|\G\|_{1,1}.
\end{equation}
This is a convex $\ell_1$-regularized least-squares problem, which can be addressed using ISTA. The method consists of a gradient-descent step applied to the quadratic term, followed by soft thresholding. Within the overall joint optimization procedure, a single ISTA iteration is performed per update. Specifically, the gradient of the quadratic term with respect to $\G$ is
\begin{equation} \label{eq:ISTA_gradient}
\nabla_{\G} \left\| \A\G\P - \Y \right\|_{\mathrm{F}}^2
= \A^{\herm} \left( \A \G \P - \Y \right) \P^{\herm},
\end{equation}
and the resulting update at iteration~$t+1$ is given by
\begin{equation} \label{eq:ISTA_update}
\G^{(t+1)} = \mathcal{S}_{\alpha\mu} \big( 
\G^{(t)} - \mu \A^{\herm} (\A \G^{(t)} \P - \Y) \P^{\herm}
\big),
\end{equation}
where $\mu > 0$ is the step size and where the threshold is the product of the step size and the regularization parameter.

\subsection{Updates of the Angles and Distances}

With $\G$ fixed, $\phib$ and $\r$ are updated to reduce the objective
\begin{equation} \label{eq:angle_range_objective}
\left\| \A(\phib, \r) \G \P - \Y \right\|_{\mathrm{F}}^2 = \|\widetilde{\E}\|_\textrm{F}^2.
\end{equation}
This is the dictionary-learning step, where the continuous angle-distance parameters are refined to better align with the actual propagation paths. To this end, we employ a Gauss-Newton scheme, in which the Hessian is approximated by the Gauss-Newton matrix \cite[p.~259]{nocedal1999numerical}. The Gauss-Newton approximation is obtained by linearizing the nonlinear model $\A(\phib, \r)$ and neglecting the additional second-order curvature term. To avoid forming and inverting the full $2Q \times 2Q$ matrix, we further adopt a diagonal approximation, updating each parameter independently using only the corresponding diagonal element. This reduces computational complexity and mainly affects the convergence speed, while retaining a descent direction under a suitable step-size selection. The derivative of \eqref{eq:angle_range_objective} with respect to $\phi_q$ and $r_q$ is given by
\begin{equation}
\label{eq:GN_gradient}
s_{\xi_q} = \frac{\partial \|\widetilde{\E}\|^2}{\partial \xi_q} =
2 \Re
\left\langle
\frac{\partial \A(\phib,\r)}{\partial \xi_q}
\G \P,
\widetilde{\E}
\right\rangle,
\quad
\xi_q \in \{\phi_q,r_q\}.
\end{equation}
The derivative of the steering matrix is given by
\begin{equation} \label{eq:steer_derivative}
    \frac{\partial \A(\phib,\r)}{\partial \xi_q} = \bigg[\0, \dots, \frac{\partial \a(\phi_q,r_q)}{\partial \xi_q}, \dots, \0 \bigg],
\end{equation}
where the elements of the nonzero vector are expressed as
\begin{align}
    \frac{\partial a_m(\phi_q,r_q)}{\partial \phi_q} &= a_m(\phi_q,r_q)j\frac{2\pi}{\lambda}\frac{r_q \delta_m d}{r_m}, \\
    \frac{\partial a_m(\phi_q,r_q)}{\partial r_q} &= -a_m(\phi_q,r_q)j\frac{2\pi}{\lambda}\left(\frac{r_q - \delta_m d \phi_q}{r_m} - 1\right), 
\end{align}
for $m = 1, \dots, M$. The diagonal entries of the Gauss-Newton matrix are defined as
\begin{equation}
\label{eq:GN_diag}
d_{\xi_q}
=
2
\left\|
\frac{\partial \A(\phib,\r)}{\partial \xi_q}
\G \P
\right\|_{\mathrm{F}}^2,
\quad
\xi_q \in \{\phi_q,r_q\},
\end{equation}
which follow from neglecting the second-order derivative term in the exact Hessian. Finally, the updates are given by
\begin{align}
\phi_q^{(t+1)} &= \Pi_{[\phi_{\textrm{min}},\,\phi_{\textrm{max}}]}
\bigg(
\phi_q^{(t)} - \delta \frac{s_{\phi_q}^{(t)}}{d_{\phi_q}^{(t)}}
\bigg), \label{eq:angle_update} \\
r_q^{(t+1)} &= \Pi_{[r_{\textrm{min}},\,r_{\textrm{max}}]}
\bigg(
r_q^{(t)} - \delta \frac{s_{r_q}^{(t)}}{d_{r_q}^{(t)}}
\bigg), \label{eq:distance_update}
\end{align}
where $\delta > 0$ is the step size and where the projections ensure that the parameters remain within physically valid limits.

\subsection{Convergence}

The objective function in \eqref{eq:DL_ISTA_objective} belongs to the class of functions considered in \cite{bolte2014proximal}, which introduces the proximal alternating linearized minimization (PALM) algorithm for solving such optimization problems. PALM alternates proximal gradient updates between subsets of optimization variables and converges to a critical point under standard regularity assumptions and sufficiently small step sizes, i.e., when the step sizes are chosen below the reciprocals of the corresponding Lipschitz constants. The proposed DL-ISTA follows a similar alternating optimization framework, but replaces the second proximal gradient update (for the angles and distances) with a projected Gauss-Newton step, which additionally incorporates local curvature information of the objective. Under appropriate damping or step-size selection, e.g., using an Armijo backtracking line search \cite{nocedal1999numerical}, the Gauss-Newton step also satisfies a sufficient decrease condition, yielding convergence behavior analogous to PALM.

\subsection{Computational Complexity}

We assume that the number of users $K$ and the pilot length $N$ are small compared with the number of antennas $M$, and therefore omit them from the asymptotic complexity expressions. We compare the computational complexity of DL-ISTA with that of P-OMP+P-SIGW \cite{cui2022channel} and CP-SBL \cite{arjas2026cross}.

\textit{\textbf{DL-ISTA.}} The gradient evaluation in~\eqref{eq:ISTA_gradient} requires $\mathcal{O}(2MQ)$ operations. Furthermore, the diagonal Gauss-Newton approximation enables each angle or distance parameter update to be computed in $\mathcal{O}(M)$ operations, yielding a total cost of $\mathcal{O}(MQ)$ across the $Q$ atoms. Consequently, the overall complexity of DL-ISTA is $\mathcal{O}\bigl(N_{\textrm{DL-ISTA}}4MQ\bigr)$, where $N_{\textrm{DL-ISTA}}$ denotes the number of DL-ISTA iterations. In our experiments, we set $Q=M$, resulting in a complexity of $\mathcal{O}\bigl(N_{\textrm{DL-ISTA}}4M^2\bigr)$.

\textit{\textbf{P-OMP+P-SIGW.}} P-OMP requires $\mathcal{O}(\widehat{L}MQ_\mathrm{p})$ operations, where $\widehat{L}$ is the number of selected atoms and $Q_\mathrm{p}$ the size of the polar-domain dictionary. For the considered setup, we have $Q_\mathrm{p}=13M$, yielding a complexity of $\mathcal{O}(13\widehat{L}M^2)$. The resulting estimate is subsequently refined using P-SIGW through $N_{\textrm{P-SIGW}}$ gradient-descent iterations, each with complexity $\mathcal{O}(M^2)$~\cite{cui2022channel}. Hence, the overall complexity of the combined P-OMP+P-SIGW pipeline is $\mathcal{O}\bigl((13\widehat{L}+N_{\textrm{P-SIGW}})M^2\bigr)$. With $N_{\textrm{DL-ISTA}} = N_{\textrm{P-SIGW}} = 50$, the two methods have comparable computational complexity at $\widehat{L} \approx 12$, whereas DL-ISTA is less computationally demanding for larger values of $\widehat{L}$. More generally, the complexities of DL-ISTA and P-OMP+P-SIGW are comparable with quadratic scaling with respect to $M$.

\textit{\textbf{CP-SBL.}} Using the same polar-domain dictionary as P-OMP, CP-SBL has per-iteration complexity $\mathcal{O}(13M^3)$ and typically converges in around $50$ iterations, making it substantially more complex than the other two approaches.

\subsection{Initialization} \label{sec:initialization}

The proposed algorithm jointly estimates the complex path gains and the continuous angle-distance parameters through alternating optimization. As a result, its convergence behavior depends strongly on the initialization of the angles and distances. Since the objective function is highly non-convex, poor initialization may lead to undesirable local minima and degraded estimation performance. Consequently, obtaining a well-distributed initial set of angle-distance pairs is critical for reliable convergence.

To mitigate this issue, we initialize the parameters using a low-discrepancy Sobol sequence \cite{sobol1967distribution}, which provides a deterministic and space-filling sampling of the parameter domain. Compared with pseudo-random sampling, Sobol sequences offer superior uniformity and thus ensure a more even coverage of the feasible region. Let $\{\s_q\}_{q=1}^Q$ denote the first $Q$ points of a two-dimensional Sobol sequence, with $\s_q = [s_{q,1}, s_{q,2}]^{\tran} \in [0,1]^2$, $\forall q \in \{1,\ldots,Q \}$. These points are mapped to the parameter domain according to
\begin{equation}
\label{eq:angle_init}
\phi_q^{(0)} = \phi_{\textrm{min}} + s_{q,1}(\phi_{\textrm{max}} - \phi_{\textrm{min}}),
\end{equation}
\begin{equation}
\label{eq:distance_init}
r_q^{(0)} = r_{\textrm{min}} + s_{q,2}(r_{\textrm{max}} - r_{\textrm{min}}).
\end{equation}
This initialization yields a well-distributed set of angle-distance pairs, providing an effective starting point for the dictionary learning and requiring significantly fewer atoms compared with the structured polar-domain grid in~\cite{cui2022channel}. The full DL-ISTA procedure is summarized in Algorithm~\ref{alg:DL-ISTA}.

\begin{algorithm}[t!]
\small
\caption{Proposed DL-ISTA for near-field channel estimation}
\label{alg:DL-ISTA}
\begin{algorithmic}[1]
\State \textbf{Input:} $\Y$, $\P$, $\phi_{\textrm{min}}$, $\phi_{\textrm{max}}$, $r_{\textrm{min}}$, $r_{\textrm{max}}$, $\alpha$, $\mu$, $\delta$
\State \textbf{Initialize:}
\For{$q = 1,\dots,Q$}
    \State Initialize $\phi_q^{(0)}$ and $r_q^{(0)}$ using \eqref{eq:angle_init}--\eqref{eq:distance_init}
\EndFor
\State Initialize $\G^{(0)} = \0$
\State $t \gets 0$
\Repeat
    \State \textbf{Update of the complex path gains (ISTA):}
    \State Update $\G^{(t+1)}$ using \eqref{eq:ISTA_update}
    
    \State \textbf{Updates of the angle and distances (dictionary learning):}
    \For{$q = 1,\dots,Q$}
        \State Update $\phi_q^{(t+1)}$ and $r_q^{(t+1)}$ using \eqref{eq:angle_update}--\eqref{eq:distance_update}
    \EndFor
    
    \State $t \gets t + 1$
\Until{convergence}
\State \textbf{Output:} $\G^t$, $\phib^t$, $\r^t$
\end{algorithmic}
\end{algorithm}

\section{Numerical Experiments}

In this section, we evaluate the performance of the proposed DL-ISTA through simulations. We first describe the simulation setup and then present and discuss the results.

\subsection{Simulation Setup}

For the performance evaluation, we vary the signal-to-noise ratio (SNR), the number of antennas, the number of pilots, and the number of propagation paths. For each parameter configuration, the channels are generated using \eqref{eq:channelmodel}--\eqref{eq:steering_vector} and the corresponding observations at the BS are obtained as in \eqref{eq:systemmodel}. The performance is assessed using the normalized mean squared error (NMSE), defined as $\textrm{NMSE} = \frac{\mathbb{E}[\|\widehat{\H} - \H\|^2]}{\mathbb{E}[\|\H\|^2]}$, where $\widehat{\H} \in \Compl^{M \times K}$ is the estimated channel matrix. The expectations are approximated via Monte Carlo simulations with 2000 independent runs. Fixing the AWGN variance to $\sigma^{2}=1$ and given $\Exp[\|\h_{k}\|^{2}] = M$ from \eqref{eq:channelmodel}--\eqref{eq:steering_vector}, the transmit power $\rho$ corresponds to the SNR.

As baselines, we consider P-OMP, P-OMP+P-SIGW \cite{cui2022channel}, the conventional ISTA without dictionary learning, and CP-SBL \cite{arjas2026cross}, all using the polar-domain dictionary defined in \cite{cui2022channel}. P-OMP selects the angle-distance pairs via orthogonal matching pursuit, whereas P-OMP+P-SIGW includes a subsequent refinement stage based on gradient descent. CP-SBL is a data-driven variant of sparse Bayesian learning that learns the sparsity-inducing weights by minimizing a randomized cross-predictive objective. The proposed DL-ISTA jointly estimates the sparse coefficients and continuous angle-distance parameters, thereby avoiding both grid mismatch effects and the additional refinement stage required by P-OMP+P-SIGW. For DL-ISTA, we set $Q=M$, resulting in a dictionary of size $M \times M$. By comparison, the size of the polar-domain dictionary in \cite{cui2022channel} depends on both the wavelength and the considered distance range. For the setup considered in this paper, the resulting dictionary size is $M \times 13M$.

In all the experiments, the Fraunhofer distance is fixed to $r_\textrm{Fraun.} = 81.92$~m, with the carrier wavelength selected accordingly. This ensures that the size of the polar-domain dictionary remains fixed at $M \times 13M$. The minimum and maximum distances are set to $r_{\textrm{min}} = \frac{r_\textrm{F}}{8}$ and $r_{\textrm{max}} = \frac{r_\textrm{F}}{2}$, respectively, while the angle cosine ranges from $\phi_{\textrm{min}} = -1$ to $\phi_{\textrm{max}} = 1$. For DL-ISTA, the regularization parameter is set to $\alpha = \frac{\sqrt{MN}}{5\sqrt{\rho L}}$, reflecting its dependence on the sparsity level, governed by the ratio $\frac{M}{L}$, and on the signal strength, determined by $\rho$ and $N$. Although $L$ is generally unknown in practice, our goal is to illustrate the potential of the method, and we thus allow $\alpha$ to depend on $L$. For P-OMP, the number of selected atoms is fixed to $3L$.

\begin{figure}[t!]
    \begin{tikzpicture}

\begin{axis}[
	width=8cm,
	height=6cm,
	xmin=-10, xmax=20,
	ymin=0.0001, ymax=0.2,
	xlabel={SNR $\rho$ [dB]},
	ylabel={NMSE},
    tick scale binop=\times,
    label style={font=\footnotesize},
    ticklabel style={/pgf/number format/fixed,font=\footnotesize},
	legend style={at={(0.98,0.98)}, anchor=north east},
	legend style={font=\scriptsize, inner sep=1pt, fill opacity=0.75, draw opacity=1, text opacity=1},
	legend cell align=left,
	grid=both,
	grid style={line width=.1pt, draw=gray!40},
    legend columns=2,
    ymode=log,
	title={},
	title style={font=\scriptsize, yshift=-2mm},
]

\addplot[thick, myBlue, mark=x, opacity=1]
table [x=Var1, y=Var3, col sep=comma] {pgfplots/results_txt/results_snr2.txt};
\addlegendentry{P-OMP};

\addplot[thick, myGreen, mark=triangle, opacity=1]
table [x=Var1, y=Var4, col sep=comma] {pgfplots/results_txt/results_snr2.txt};
\addlegendentry{P-OMP+P-SIGW};

\addplot[thick, myOrange, mark=square, opacity=1]
table [x=Var1, y=Var5, col sep=comma] {pgfplots/results_txt/results_snr2.txt};
\addlegendentry{ISTA};

\addplot[thick, myPurple, mark=star, opacity=1]
table [x=Var1, y=Var6, col sep=comma] {pgfplots/results_txt/results_snr2.txt};
\addlegendentry{CP-SBL};

\addplot[very thick, myRed, mark=o]
table [x=Var1, y=Var2, col sep=comma] {pgfplots/results_txt/results_snr2.txt};
\addlegendentry{DL-ISTA};

\end{axis}

\end{tikzpicture}
    \caption{NMSE versus SNR, with $M = 256$, $N = 20$, $K = 5$, and $L = 5$.}
    \label{fig:nmse_snr}
\end{figure}
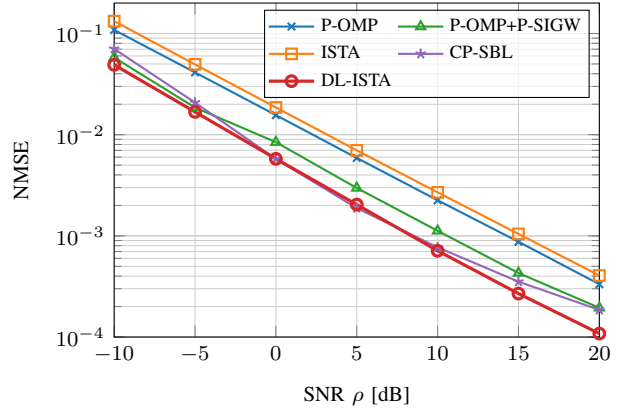

\subsection{Results}

Fig.~\ref{fig:nmse_snr} plots the NMSE as a function of the SNR $\rho$. All the methods exhibit approximately linear decrease in NMSE with increasing SNR. DL-ISTA consistently achieves the lowest NMSE across the entire SNR range. While CP-SBL attains nearly identical performance, a slight saturation effect can be observed at high SNR. Importantly, this comparable estimation accuracy is achieved by DL-ISTA with a substantially lower computational cost. In contrast, P-OMP+P-SIGW exhibits a clear performance gap throughout the considered SNR range, with DL-ISTA achieving approximately $40\%$ lower NMSE at $\rho = 20$~dB. The performance gap between DL-ISTA and the conventional ISTA highlights the importance of the dictionary-learning step during the estimation process.

\begin{figure}[t!]
    \begin{tikzpicture}

\begin{axis}[
	width=8cm,
	height=6cm,
	xmin=64, xmax=256,
	ymin=0.0005, ymax=0.0055,
	xlabel={Number of antennas $M$},
	ylabel={NMSE},
	xtick={64,96,...,256},
	ytick={0.001,0.002,...,0.005},
    tick scale binop=\times,
    label style={font=\footnotesize},
    ticklabel style={/pgf/number format/fixed,font=\footnotesize},
	legend style={at={(0.98,0.98)}, anchor=north east},
	legend style={font=\scriptsize, inner sep=1pt, fill opacity=0.75, draw opacity=1, text opacity=1},
	legend cell align=left,
	grid=both,
	grid style={line width=.1pt, draw=gray!40},
    legend columns=2,
	title={},
	title style={font=\scriptsize, yshift=-2mm},
]

\addplot[thick, myBlue, mark=x, opacity=1]
table [x=Var1, y=Var3, col sep=comma] {pgfplots/results_txt/results_M2.txt};
\addlegendentry{P-OMP};

\addplot[thick, myGreen, mark=triangle, opacity=1]
table [x=Var1, y=Var4, col sep=comma] {pgfplots/results_txt/results_M2.txt};
\addlegendentry{P-OMP+P-SIGW};

\addplot[thick, myOrange, mark=square, opacity=1]
table [x=Var1, y=Var5, col sep=comma] {pgfplots/results_txt/results_M2.txt};
\addlegendentry{ISTA};

\addplot[thick, myPurple, mark=star, opacity=1]
table [x=Var1, y=Var6, col sep=comma] {pgfplots/results_txt/results_M2.txt};
\addlegendentry{CP-SBL};

\addplot[very thick, myRed, mark=o]
table [x=Var1, y=Var2, col sep=comma] {pgfplots/results_txt/results_M2.txt};
\addlegendentry{DL-ISTA};

\end{axis}

\end{tikzpicture}
    \caption{NMSE versus $M$, with SNR = $10$ dB, $N = 20$, $K = 5$, and $L = 5$.}
    \label{fig:nmse_M} \vspace{-3mm}
\end{figure}
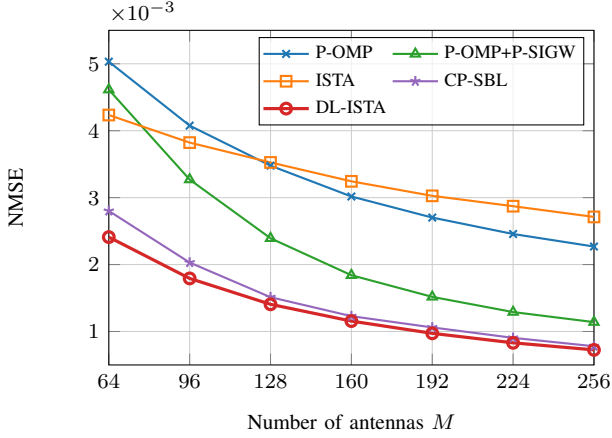

\begin{figure}[t!]
    \begin{tikzpicture}

\begin{axis}[
	width=8cm,
	height=6cm,
	xmin=5, xmax=95,
	ymin=0.0001, ymax=0.01,
	xlabel={Pilot length $N$},
	ylabel={NMSE},
	xtick={5,15,...,95},
    tick scale binop=\times,
    label style={font=\footnotesize},
    ticklabel style={/pgf/number format/fixed,font=\footnotesize},
	legend style={at={(0.98,0.98)}, anchor=north east},
	legend style={font=\scriptsize, inner sep=1pt, fill opacity=0.75, draw opacity=1, text opacity=1},
	legend cell align=left,
	grid=both,
	grid style={line width=.1pt, draw=gray!40},
    legend columns=2,
    ymode=log,
	title={},
	title style={font=\scriptsize, yshift=-2mm},
]

\addplot[thick, myBlue, mark=x, opacity=1]
table [x=Var1, y=Var3, col sep=comma] {pgfplots/results_txt/results_N2.txt};
\addlegendentry{P-OMP};

\addplot[thick, myGreen, mark=triangle, opacity=1]
table [x=Var1, y=Var4, col sep=comma] {pgfplots/results_txt/results_N2.txt};
\addlegendentry{P-OMP+P-SIGW};

\addplot[thick, myOrange, mark=square, opacity=1]
table [x=Var1, y=Var5, col sep=comma] {pgfplots/results_txt/results_N2.txt};
\addlegendentry{ISTA};

\addplot[thick, myPurple, mark=star, opacity=1]
table [x=Var1, y=Var6, col sep=comma] {pgfplots/results_txt/results_N2.txt};
\addlegendentry{CP-SBL};

\addplot[very thick, myRed, mark=o]
table [x=Var1, y=Var2, col sep=comma] {pgfplots/results_txt/results_N2.txt};
\addlegendentry{DL-ISTA};

\end{axis}

\end{tikzpicture}
    \caption{NMSE versus $N$, with SNR = $10$ dB, $M = 256$, $K = 5$, and $L = 5$.}
    \label{fig:nmse_N}
\end{figure}
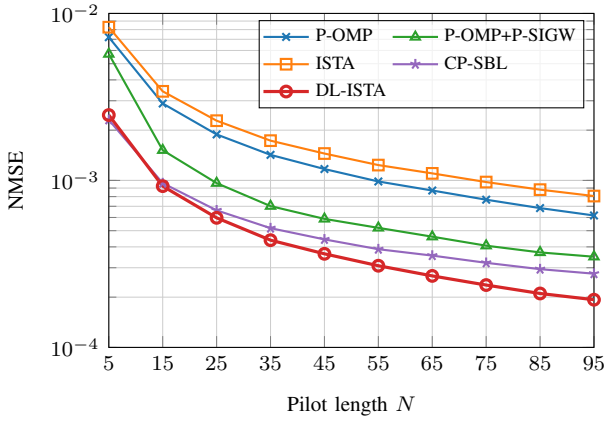

Fig.~\ref{fig:nmse_M} shows the NMSE as a function of the number of antennas $M$. DL-ISTA and CP-SBL maintain nearly identical performance across the entire range of antenna configurations, further demonstrating that the proposed method achieves state-of-the-art estimation accuracy without the computational burden associated with CP-SBL. Although the performance gap with respect to P-OMP+P-SIGW gradually decreases with $M$, DL-ISTA consistently provides the lowest NMSE. Fig.~\ref{fig:nmse_N} illustrates the NMSE as a function of the pilot length $N$. For small $N$, DL-ISTA and CP-SBL are quite closely matched but the gap widens in favor of DL-ISTA when $N$ increases. P-OMP+P-SIGW consistently yields higher NMSE values.

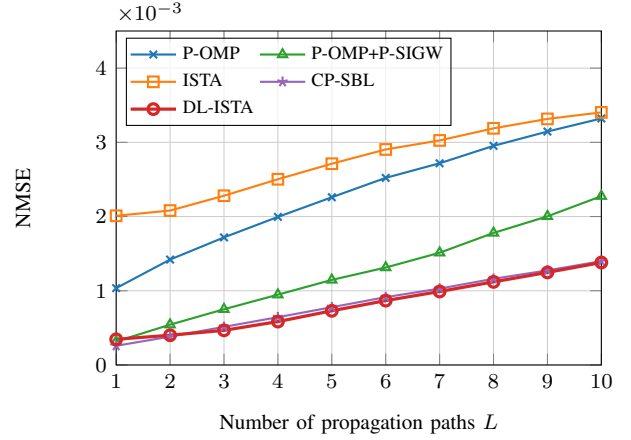
\begin{figure}[t!]
    \begin{tikzpicture}

\begin{axis}[
	width=8cm,
	height=6cm,
	xmin=1, xmax=10,
	ymin=0, ymax=0.0045,
	xlabel={Number of propagation paths $L$},
	ylabel={NMSE},
	xtick={1,2,3,...,10},
	ytick={0,0.001,...,0.004},
    tick scale binop=\times,
    label style={font=\footnotesize},
    ticklabel style={/pgf/number format/fixed,font=\footnotesize},
	legend style={at={(0.02,0.98)}, anchor=north west},
	legend style={font=\scriptsize, inner sep=1pt, fill opacity=0.75, draw opacity=1, text opacity=1},
	legend cell align=left,
	grid=both,
	grid style={line width=.1pt, draw=gray!40},
    legend columns=2,
	title={},
	title style={font=\scriptsize, yshift=-2mm},
]

\addplot[thick, myBlue, mark=x, opacity=1]
table [x=Var1, y=Var3, col sep=comma] {pgfplots/results_txt/results_L2.txt};
\addlegendentry{P-OMP};

\addplot[thick, myGreen, mark=triangle, opacity=1]
table [x=Var1, y=Var4, col sep=comma] {pgfplots/results_txt/results_L2.txt};
\addlegendentry{P-OMP+P-SIGW};

\addplot[thick, myOrange, mark=square, opacity=1]
table [x=Var1, y=Var5, col sep=comma] {pgfplots/results_txt/results_L2.txt};
\addlegendentry{ISTA};

\addplot[thick, myPurple, mark=star, opacity=1]
table [x=Var1, y=Var6, col sep=comma] {pgfplots/results_txt/results_L2.txt};
\addlegendentry{CP-SBL};

\addplot[very thick, myRed, mark=o]
table [x=Var1, y=Var2, col sep=comma] {pgfplots/results_txt/results_L2.txt};
\addlegendentry{DL-ISTA};

\end{axis}

\end{tikzpicture}
    \caption{NMSE versus $L$, with SNR = $10$ dB, $M = 256$, $N = 20$, and $K = 5$.}
    \label{fig:nmse_L}
\end{figure}

Lastly, Fig.~\ref{fig:nmse_L} plots the NMSE as a function of the number of propagation paths $L$. The channel estimation becomes increasingly challenging as $L$ increases, leading to higher NMSE for all the methods. For the single-path case ($L=1$), DL-ISTA and P-OMP+P-SIGW achieve similar performance. However, as the $L$ increases, the performance advantage of DL-ISTA becomes progressively more pronounced, suggesting that the proposed method is particularly effective in more complex propagation environments. Throughout the entire range of $L$, DL-ISTA remains virtually indistinguishable from CP-SBL in terms of NMSE performance, despite the substantially higher computational complexity of the latter.

\section{Conclusion}

We proposed DL-ISTA, a method for joint near-field dictionary learning and channel estimation that jointly estimates the sparse channel coefficients and the continuous angle-distance parameters, thereby avoiding discretization errors from fixed grids. Initialization based on Sobol sequences was employed to uniformly cover the parameter space and promote robust convergence. Numerical results demonstrated that DL-ISTA consistently achieves the lowest NMSE among the considered methods, also outperforming CP-SBL while requiring approximately one order of magnitude lower computational complexity. Hence, DL-ISTA offers an attractive balance between estimation accuracy and computational efficiency for near-field channel estimation in XL-MIMO. Future work will investigate automatic regularization parameter selection to improve the practical applicability and robustness of DL-ISTA.

\bibliographystyle{IEEEtran}
\bibliography{refs_abbr,refs}

\end{document}